\documentclass[aps,pre,twocolumn,showpacs,amsmath,amssymb]{revtex4}
\usepackage{graphicx}% Include figure files
\usepackage{dcolumn}% Align table columns on decimal point
\usepackage{bm}% bold math

\begin{document}

\preprint{APS/123-QED}

\title[Ising Ferromagnet with Annealed Vacancies]{A Blume-Capel Ising Ferromagnet with Annealed Vacancies on a Hierarchical Lattice}

\author{Daniel P. Snowman}
%\email{dsnowman@ric.edu}
%\thanks{}

\affiliation{Department of Physical Sciences, Rhode Island College, \\Providence, Rhode Island 02908}

\date{\today}% It is always \today, today,

\begin{abstract}
A dilute Ising ferromagnet is considered in this study using renormalization group techniques with a hierarchical lattice.  A series of phase diagrams have been produced that probe the effects of varying the temperature and concentration of nonmagnetic impurities.  Each phase diagram corresponds to a different strength for the internal coupling coefficients on our lattice.  Phases have been interpreted and critical exponents calculated for the higher order transitions.  
\end{abstract}

\pacs{5.70.Fh, 64.60.-i, k75.10.Nr, 5.50.+q}% PACS, the Physics and Astronomy
\maketitle

\section{Introduction}
The Blume-Capel model is a spin-1 Ising model with a Hamiltonian having bilinear ($J_{ij}$ and crystal-field ($\Delta_{ij}$) interactions. 

\begin{eqnarray}
- \beta H = \sum_{\langle{ij}\rangle} J_{ij} s_i s_j  - \sum_{\langle {ij}\rangle}\Delta_{ij}(s_{i}^2 + s_{j}^2) 
\nonumber \\
 \text{with} \qquad s_i = 0, \pm 1 \qquad \qquad \qquad \qquad
\end{eqnarray}
 
The concentration of nonmagnetic impurities ($s_i=0$), or annealed vacancies, in our system is directly related to the crystal field interaction ($\sim \Delta/J$).  While the traditional Ising interaction ($J_{ij})$ primarily effects magnetic ordering.  Systems driven by fluctuations in both magnetization and density are particularly well suited for study using this model.  Each contribution to the Hamiltonian in Eq. 1 involves a summation over nearest-neighbor $\langle{ij}\rangle$ pairs of our sites on our lattice unit structure including the crystal-field interaction term.  Spin-1 Ising models, with density as an additional degree of freedom, have been used to probe, investigate and further understand a range of very complex systems.  Included amongst these is the superfluid transitions in $He^{3}-He^{4}$ mixtures ~\cite{Blume:PRA71}, structural glasses ~\cite{Kirkpatrick:PRB97},  binary fluids, materials with mobile defects, semiconductor alloys ~\cite{Newman:PRB83}, microemulsions~\cite{Schick:PRB86}, frustrated percolation systems ~\cite{ Coniglio:JPF93} and aerogels~\cite{Maritan:PRL92}.

Many different types of competing interactions have been the focus of previous studies using the Blume-Emery-Griffiths model in conjunction with mean-field methods ~\cite{Hoston:PRL91, Sellitto:JPF97, Snowman:PhD95, McKay:JAP84} and/or renormalization-group techniques ~\cite{Berker:PRB76, McKay:PRL82, Branco:lanl99, Branco:lanl97, Snowman:JMMM07, Snowman:JMMM08, Snowman:PRE08}.  In many studies, it is found that competition between underlying microscopic interactions can drastically alter phase diagrams and associated criticality in various Ising systems.

Competing bilinear interactions ~\cite{McKay:PRL82} in a spin-1/2 Ising model, competing bilinear interactions in a BEG system ~\cite{Snowman:JMMM08}, competing biquadratic interactions in a dilute Ising ferromagnet ~\cite{Snowman:PRE08}, and simultaneous competition between crystal-field and biquadratic interactions in a BEG ferromagnet ~\cite{Snowman:JMMM07} have all been investigated using  renormalization-group techniques in concert with hierachical lattices.

Other studies have considered the effects of quenched random bonds ~\cite{ Falicov:PRL96} and quenched random fields ~\cite{ Kabak:PRL99} upon the criticality and phase diagrams in BEG systems.  Branco et al. ~\cite{Branco:lanl99,Branco:lanl97, Branco:lanl98} have considered the effects of random crystal fields using real-space renormalization-group methods and mean-field approximations for both Blume-Emery-Griffiths and Blume-Capel hamiltonians, respectively.  

The current study complements these earlier works as it considers a dilute Ising ferromagnet and the effect of varying the concentration of nonmagnetic impurities in the system.  A series of phase diagrams have been produced while varying the relative strength of internal coupling coefficients in the system.  For the model considered (dilute Ising on a hierarchical lattice) and techniques employed (renormalization group theory) the existing literature lacked data, thus, the purpose of the results and discussion presented here.

\section{Renormalization Group and Hierarchical Lattices and }
	
In general, infinite hierarchical lattices are constructed from a basic unit, or generator, by repeatedly replacing each bond by the basic unit itself.  A generic hierarchical lattice is constructed and illustrated in Figures 1.  Figure 2 illustrates the construction of the more complex hierarchical lattice ~\cite{Berker:JPC79,Kauffman:PRB81} used for the present study.  The increased complexity due to changes in the internal connectivity of the basic generating unit structure.  Since these specialized lattices yield exact renormalization group recursion relations, hierarchical lattices are very attractive to use as model systems.  Thus, critical scaling exponents and phase diagrams can be calculated very precisely.  Several previous studies have employed hierarchical lattices to effectively probe spin glass ~\cite{Snowman:JMMM08, Migliorini:PRB98}, frustrated ~\cite{McKay:PRL82, Snowman:JMMM07,Snowman:PRE08}, random-bond ~\cite{ Andelman:PRB84}, random-field ~\cite{Falicov:PRB95}, directed-path ~\cite{daSilv:PRL04} and dynamic scaling ~\cite{Stinchcombe:JPA86} systems.  

\begin{figure}
\begin{center}
\leavevmode
\includegraphics[scale=0.5]{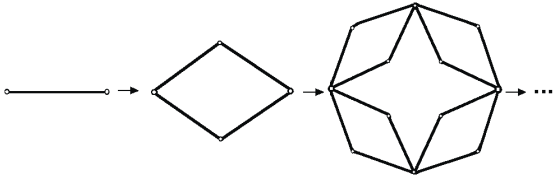} 
\end{center}
\caption{An infinite hierarchical lattice generated from a basic unit (Berker and Ostlund ~\cite{Berker:JPC79}).}
\label{fig:Figure 1}
\end{figure}

Renormalization reverses the construction process for the infinite hierarchical lattice as internal degrees of freedom are eliminated by summing over all configurations of the innermost spin sites (represented by solid black dots in Figures 2a and 2b, represented by {$s_i, s_j$} in Equation 5).

\begin{figure}
\begin{center}
\leavevmode
\includegraphics[scale=0.5]{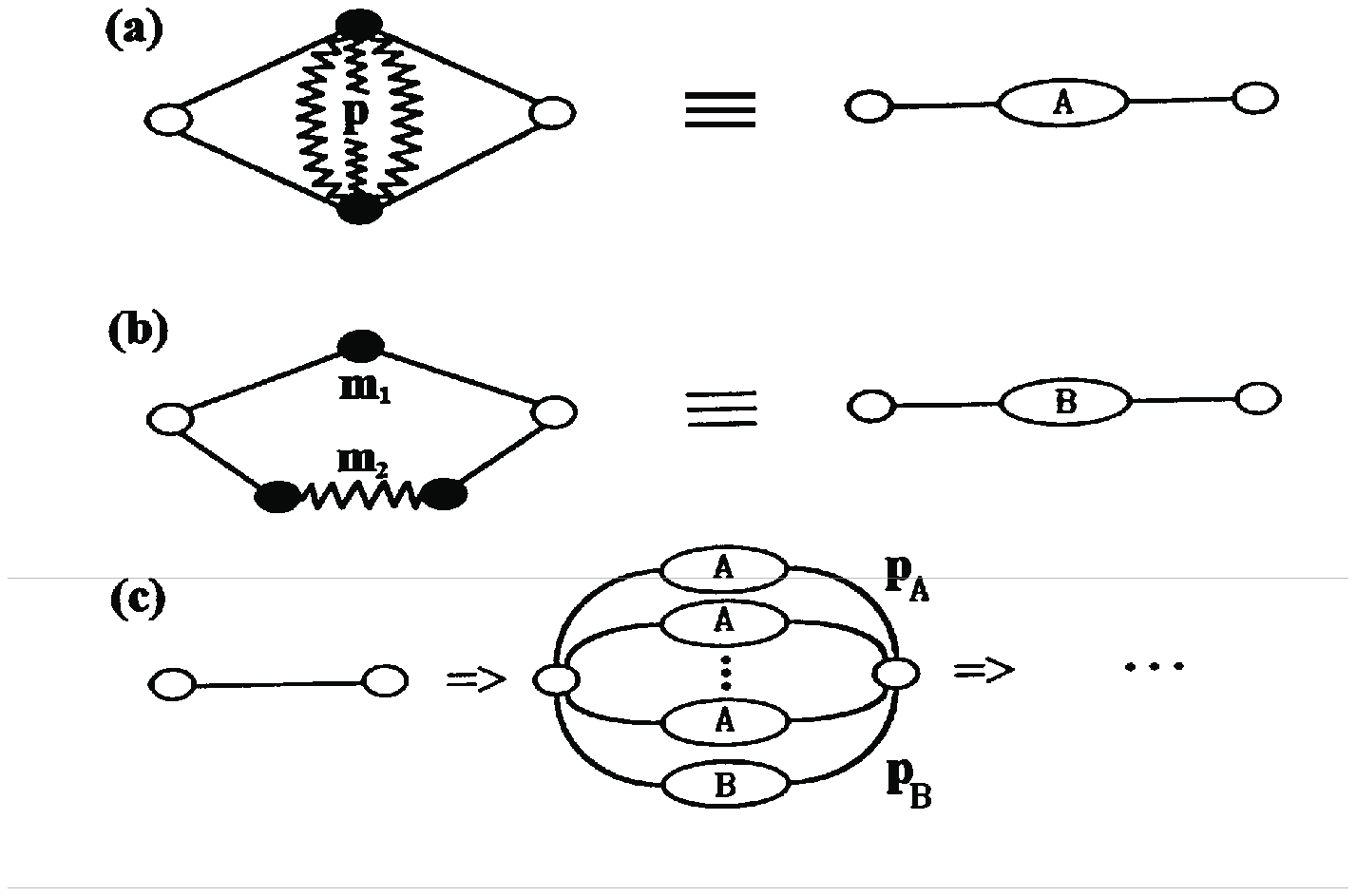} 
\end{center}
\caption{Construction of the hierarchical lattice used in this study.  (Reprinted from Journal of Magnetism and Magnetic Materials, 314, 69-74 (2007) D. P. Snowman,  with permission from Elsevier)}
\label{fig:Figure 2}
\end{figure}

Renormalization-group relations, relating the coupling coefficients at the two length scales, are developed by requiring the partition function to remain unchanged with each rescaling.  The new effective coupling coefficients $J'$, and $\Delta'$ are separated by a distance $l'$ which is $b$ lattice constants in the original system, where b is the length rescaling factor of the renormalization-group transformation.  Under renormalization, biquadratic interactions arise that must also be considered.  Thus, in general,

\begin{eqnarray}
\zeta_{l'}(J',K',\Delta ')= \zeta_{l}(J,K,\Delta) \\
\nonumber\\
\text{with} \; l' = b l \qquad \qquad 
\end{eqnarray}

\begin{eqnarray}
\zeta_l = \sum_{s} \exp[-\beta H] = \sum_{} R_l (s_i,s_j)\\
\text{with}\; R_l(s_i,s_j) = \sum_{<ij>}\exp[-\beta H]
\end{eqnarray}

\begin{eqnarray}
\zeta_{l'} = \sum_{s_i^{'},s_j^{'}} \exp[-\beta H'] = \sum_{s_i^{'}, s_j^{'}} R_l (s_i^{'},s_j^{'}) \\
\text{with} \; R_l(s_i^{'},s_j^{'}) = \sum_{s_i^{'}, s_j^{'}}\exp[J's_i s_j \nonumber \\ \; + K' s_i^2 s_j^2  - \Delta ' (s_i^2+s_j^2) + \widetilde{G}']
\end{eqnarray}
  
where $\widetilde{G}'$ is a constant used to calculate the free energy.   

The derivation for the renormalzation-group transformation for each coupling coefficient is calculated by equating individual contributions, $R_{l}(s_i,s_j)$ and $R_{l'}(s_i,s_j)$, to the summation for the partition function at each length scale.  These contributions to the partition function, $R_{l}(s_i,s_j)$ and $R_{l'}(s_i,s_j)$, correspond to the same fixed configuration of end spins, ${s_i, s_j}$, at the two different length scales, $l$ and $l'$.  From the resulting relationships, we algebraically derive relations between interaction strengths at the two length scales, $l$ and $l'$: $J' (J, K, \Delta)$, $J' (J, K, \Delta)$, $\Delta' (J, \Delta)$.  The reader is directed to Section 5 for a derivation of these relations.

Phase diagrams are mapped and the order of each transition determined using these recursion relations in conjunction with the initial values of $J$ and $\Delta$, and the resulting flow and sink(s) of the renormalization-group trajectories.

\begin{eqnarray}  
J' = R_J (J, K, \Delta)	\\
\nonumber\\
K' = R_K (J, K, \Delta)	\\	
\nonumber\\  			 
\Delta ' = R_{\Delta} (J, K, \Delta) \\
\nonumber
\end{eqnarray}

With each phase there exists an associated phase sink, see Table I, at which the values of the interactions $(J, K, \Delta)$ have reached a fixed point denoted by $(J^{*}, K^{*},\Delta^{*})$.  At these fixed points the system is scale invariant and as a consequence renormalization does not effect the properties of the system as the length scale is increased by a factor of $b$.  That is, the fixed points must satisfy the recursion relations such that 

\begin{eqnarray}
J^{*} = R_J(J^{*}, K^{*}, \Delta^{*}) \\
\nonumber\\
K^{*} = R_K(J^{*}, K^{*}, \Delta^{*})\\
\nonumber\\				
\Delta^{*} = R_{\Delta}(J^{*}, K^{*}, \Delta^{*})\\
\nonumber
\end{eqnarray}

The hierarchical lattice used in the current study is generated from a unit structure with two types of components (see Fig. 2a and b), similar to references ~\cite{McKay:PRL82, McKay:JAP82,Snowman:JMMM07,Snowman:JMMM08,Snowman:PRE08}.  One type of component, type A, is distinguished by a cross-link feature of strength p thus allowing internal spins to interact via nearest neighbor interaction $J,\Delta$.  The limiting case of $p=0$ corresponds to the hierarchical equivalent ~\cite{Berker:JPC79} of the Migdal-Kadanoff ~\cite{Migdal75, Kadanoff76} decimation-bond moving scheme in two dimensions.    

In addition, a second type of component, type B as shown in Figure 2b, allows the end spins to interact via two different connecting paths consisting of $m_1$ and $m_2$ pairs of spins, respectively.  The relative number of each type of component, used in our basic generating unit, is controlled via two parameters $p_A$ and $p_B$.  
 
\section{Phase Transition Characterization}
The free energy density (dimensionless free energy per bond), $f$, can be expressed as
  
\begin{equation}
f = {- \frac{\beta F}{N_{b}}} = \sum_{n=1}^{\infty}b^{-nd}{G'^{(n)}}(J^{(n-1)},K^{(n-1)},\Delta^{(n-1)}  
\end{equation}

where $F$ is the Helmhotz free energy and $N_b$ denotes the total number of bonds in the system.  The free energy density consists of a sum, over all iterations of the renormalization-group transformation, of the contributions $G'^{(n)}$ to the free energy density due to the degrees of freedom removed during each transformation.  Each renormalization-group transformation reduces the length scale of the system by a factor of $b$ and the number of spins by a factor of $b^d$.  

Numerically differentiating the free energy density allows us to calculate densities, magnetizations, and nearest neighbor correlations.  For example, the magnetization, $m \equiv \frac{M}{N_s} =\frac{N_b}{N_s}\frac{\delta f}{\delta H}$, can be calculated by numerically measuring the shift in the free energy density with a small perturbation in the magnetic field, where $N_s$ is the number of sites.  Similarly, the density can be calculated by differentiating the free energy density with respect to the crystal field coefficient, $\rho \equiv \frac{N_b}{N_s}\frac{\delta f}{\delta \Delta}$ .  Nearest neighbor correlations of the bilinear, $\langle s_i s_j \rangle = \frac{N_b}{N_s}\frac{\delta f}{\delta J}$, exchange interactions are also valuable when interpreting the phases and characterizing transitions.  

The order of each transition is characterized using the four thermodynamic quantities discussed above with first order transitions being signaled with discontinuities in any one of these order parameters.  Note, second order or continuous transitions exhibit no such discontinuities.  In addition, critical scaling exponents can be calculated at critical transitions since we have exact recursion relations.  

\section{Recursion Relations}
For each fixed end-spin configuration, we equate the contributions to the partition function from the two length scales, allowing us to write the following equalities for the type A structure shown in Figure 2a, for the general case with nonzero $K$.  

\begin{widetext}
\begin{eqnarray}
R_l[1,1] & = & \exp[-4 \Delta]+2 \exp[-2 J+2 K-\Delta (6+p)]
\nonumber \\
& & + 2 \exp[2 J+2 K-\Delta (6+p)]+
  \exp[J (-4+p)+K (4+p)-\Delta (8+2 p)]
\nonumber \\
& & + 2 \exp[-J p+K (4+p)-\Delta (8+2 p)]+
  \exp[J (4+p)+K (4+p)-\Delta (8+2 p)] 
\nonumber \\
& & = \exp[J' + K' - 2 \Delta' + \tilde{G}] = R_{l'}[1,1],
\end{eqnarray}

\begin{eqnarray}
R_l[1,0] & = & \exp[-2 \Delta]+2 \exp[-J+K-\Delta (4+p)]+2 \exp[J+K-\Delta (4+p)]
\nonumber \\
& & + \exp[J (-2+p)+K (2+p)-\Delta (6+2 p)]+2 \exp[-J p+K (2+p)-\Delta (6+2 p)]
\nonumber \\
& & +\exp[J (2+p)+K (2+p)-\Delta (6+2 p)] = \exp[-\Delta' + \tilde{G}] = R_{l'}[1,0],
\end{eqnarray}

\begin{eqnarray}
R_l[1,-1] & = & \exp[-4 \Delta]+4 exp[2 K-\Delta (6+p)]
\nonumber \\
& & + 2 \exp[-J p+K (4+p)-\Delta (8+2 p)]+2 
    \exp[J p+K (4+p)-\Delta (8+2 p)]
\nonumber \\
& & = \exp[- J' + K' - 2\Delta' + \tilde{G}] = R_{l'}[1,-1],
\end{eqnarray}

\begin{eqnarray}
R_l[0,0] & = & 1 + 4 \exp[-\Delta (2+p)]+2 \exp[-J p+K p-\Delta (4+2 p)]
\nonumber \\
& & + 2 \exp[J p+K p-\Delta (4+2 p)] = \exp[\tilde{G}] = R_{l'}[0,0],
\end{eqnarray}
\end{widetext}

The desired renormalization-group transformations, relating the coupling coefficients between the two length scales for the type A unit structure, can be derived by algebraically manipulating the relationships above (Eqs. 15-18).

\begin{eqnarray}
J_A^{'} = \frac{1}{2}\log{\frac{R_{l'}(1,1)}{R_{l'}(1,-1)}}\qquad\qquad\qquad\\
\nonumber\\
K_A^{'} = \frac{1}{2}\log{\frac{R_{l'}(1,1) R_{l'}(1,-1) R_{l'}^2(0,0)}{R_{l'}^4(1,0)}}\\
\nonumber\\
\Delta_A^{'} = \log{\frac{R_{l'}(0,0)}{R_{l'}(1,0)}}\qquad\qquad\qquad\\
\nonumber\\
\widetilde{G}_A^{'} = \log{R_{l'}(0,0)}\qquad\qquad\qquad\\
\nonumber
\end{eqnarray}

The recursion relations for the less complex (type B) unit structures have the same form as in Eqs. 19-22, but the expressions (Eqs. 15-18) for the corresponding $R_l(s_i,s_j)$ differ.  Combining the contributions from both types of unit structures (type A and type B as shown in Fig. 2), the renormalization relationships become

\begin{eqnarray}
J^{'} = p_A J_A^{'} + p_B J_B^{'}\\
\nonumber\\
K^{'} = p_A K_A^{'} + p_B K_B^{'}\\
\nonumber\\
\Delta{'} = p_A \Delta_A^{'} + p_B \Delta_B^{'}\\
\nonumber
\end{eqnarray}

The exact nature of the renormalization-group relations above, allow us to calculate critical exponents by linearizing the recursion relations in the vicinity of the critical transition under investigation.  That is, 

\begin{eqnarray}
J^{'}-J^{*} & = & T_{JJ} (J-J^{*})+T_{JK} (K-K^{*})\nonumber\\
& & + T_{J{\Delta}}(\Delta-\Delta^{*}),
\end{eqnarray}

\begin{eqnarray}
K^{'}-K^{*} & = & T_{KJ} (J-J^{*})+T_{KK} (K-K^{*})\nonumber\\
& & + T_{K{\Delta}}(\Delta-\Delta^{*}),
\end{eqnarray}

\begin{eqnarray}
\Delta^{'}-\Delta^{*} & = & T_{{\Delta}J}(J-J^*) + T_{{\Delta}K}(K-K^*)\nonumber\\
& & + T_{{\Delta}{\Delta}}(\Delta-\Delta^{*}),
\end{eqnarray}

where  $T_{JJ} = \frac{\delta J'}{\delta J}$, $T_{{\Delta}J} = \frac{\delta {\Delta}'}{\delta J}$, etc. and are evaluated at the fixed point in question.  A recursion matrix, with elements $T_{XY}$ and eigenvalues of the form

\begin{equation}
\Lambda_l = b^{y_l}
\end{equation}

can be used to represent the critical relations in Eqs. 26-28.  Here, $b$ is the length rescaling factor (in our case $b=2$) and $y_l$ represents the corresponding critical exponent for the $l^{th}$ eigenvalue.  Critical scaling exponents have also been calculated for nonzero perturbations of the odd sector components H and L.

\section{Results}

Our results below probe the effects upon ordering of varying the temperature (1/J) and vacancy concentration ($~{\Delta}/J$) present in the system.  A series of phase diagrams are produced, each corresponding to a different level of internal interactions, tuned via the parameter p.  Exhaustive analysis of renormalization-group trajectories and corresponding sinks yields three phases:  a ferromagnetic, a dense paramagnetic and a dilute paramagnetic phase.  Each phase shares a common sink or basin of attraction in parameter space, as detailed in Table I.  

\begin{table}
\begin{tabular}{|l|l|l|}
\hline  Phase & Sink & Characteristics \\ 
\hline  Dense Paramagnetic & $J \rightarrow 0$ & Low concentration of\\ 
  &  $\Delta \rightarrow -\infty$ & nonmagnetic impurities\\  
\hline  Ferromagnetic& $J \rightarrow +\infty$ & Magnetically\\ 
  & $\Delta \rightarrow -\infty$ & ordered\\ 
\hline  Dilute Paramagnetic& $J \rightarrow 0$ & High concentration of\\ 
  &  $\Delta \rightarrow +\infty$ & nonmagnetic impurities\\ 
\hline 
\end{tabular}
\vspace{3mm}
\caption{Phases and Corresponding Sinks}
\end{table}

The dense and dilute paramagnetic phases are distinguished from one another via the flow of the crystal-field interaction term.  A flow to -$\infty$ corresponding to a dense population of magnetic species, whereas a flow to +$\infty$ corresponds to a system to dilute for magnetic ordering to occur.  For each paramagnetic phase, the bilinear (J) interaction flows to zero indicating no magnetic ordering.  In addition, a ferromagnetic phase arises with a renormalization group flow with the bilinear interaction flowing to +$\infty$ and the crystal-field interaction flowing to -$\infty$.  The ferromagnetic phase can be reached by decreasing the temperature, or, by decreasing the concentration of nonmagnetic impurities.

\begin{figure}
\begin{center}
\leavevmode
\includegraphics[scale=0.5]{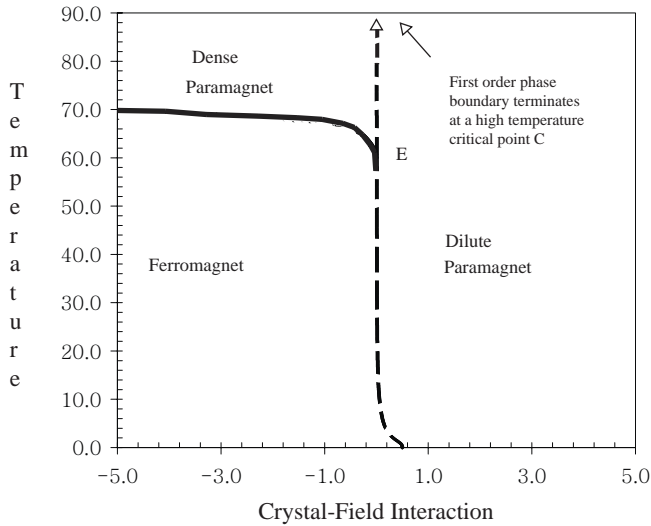} 
\end{center}
\caption{Phase diagram with connectivity values
($p, m_1, m_2, p_A, p_B$) = (1, 8, 9, 40, 1), showing different basins of attraction and associated phases with critical endpoint (E) and critical point (C). Solid lines represent second-order transitions, whereas dashed lines represent first-order transitions.}
\label{fig:Figure 3}
\end{figure}

The results presented below were generated while considering various planes of parameter space with different degrees of internal connectivity.  The parameter p allows us to vary this connectivity as we investigate the effects upon the underlying phase diagrams of changes in temperature (1/J) and crystal-field interaction ($~{\Delta}/J$).  In the first plane we consider the plane with p=1 and we find two paramagnetic phases (dense and dilute) and a magnetically ordered ferromagnetic phase.  The ferromagnetic phase occurs at intermediate and lower temperatures for values of the crystal-field interaction (${\Delta}/J$) that correspond to a lower concentration of nonmagnetic impurities on our lattice.

\begin{figure}
\begin{center}
\leavevmode
\includegraphics[scale=0.5]{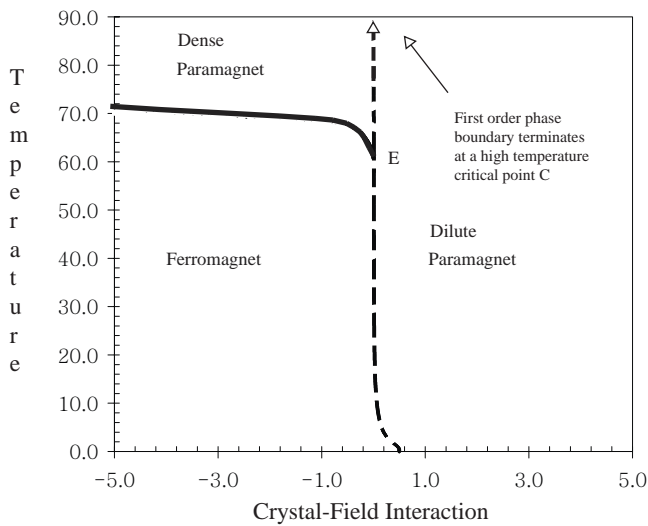} 
\end{center}
\caption{Phase diagram with connectivity values
($p, m_1, m_2, p_A, p_B$) = (2, 8, 9, 40, 1), showing different basins of attraction and associated phases with critical endpoint (E) and critical point (C).  Solid lines represent second-order transitions, whereas dashed lines represent first-order transitions.}
\label{fig:Figure 4}
\end{figure}

The ferromagnetic phase disorders at high temperature via a second order transitions to a dense paramagnetic state.  In this state, the concentration of occupied sites is such that magnetic order is possible, however, the temperature is too great for long-range magnetic order to propagate.  The dense paramagnetic phase is present only at high temperatures and is separated from its dilute paramagnetic counterpart via a line of first order phase boundary that terminates at a high temperature critical point C.  At temperatures above this critical point, it is possible to drive the system from one paramagnetic phase to the other sans transition as in the standard liquid-gas phase diagram.  The line of criticality separating the ferromagnetic and dense paramagnetic states terminates at a critical endpoint E upon the intersection with the line of first order transitions.

The dilute paramagnetic state appears at those values of the crystal-field interaction that correspond to a concentration of nonmagnetic impurities that exceeds a certain percolation threshold necessary in order for long-range magnetic order to propagate. 

\begin{figure}
\begin{center}
\leavevmode
\includegraphics[scale=0.5]{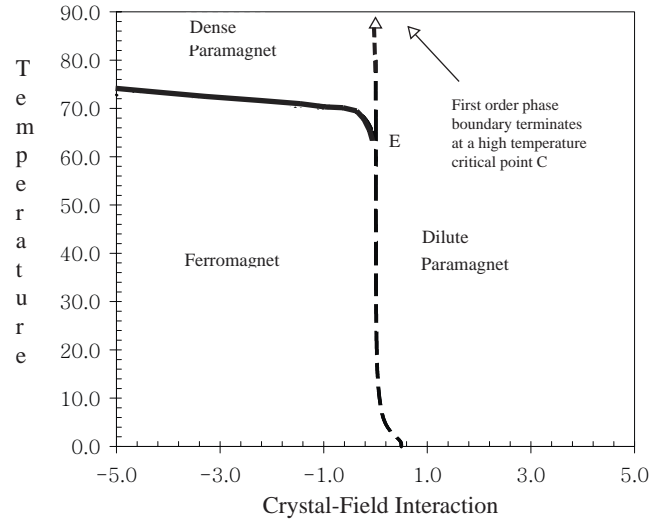} 
\end{center}
\caption{Phase diagram with connectivity values
($p, m_1, m_2, p_A, p_B$) = (4, 8, 9, 40, 1), showing different basins of attraction and associated phases with critical endpoint (E).  Solid lines represent second-order transitions, whereas dashed lines represent first-order transitions.}
\label{fig:Figure 5}
\end{figure}

An increase in the strength of the internal coupling coefficients to p=2 reveals an underlying phase diagram that is very similar to the p=1 case.  The location of the first order phase boundary remains unchanged, and nearly vertical at ${\Delta}/J = 0$.  Thus,  our system can be forced to transition between the two paramagnetic states at high temperature only through a change in the crystal-field interactions present.  That is, we must change the concentration of occupied sites to traverse this first-order phase boundary.  Also, the line of criticality separating the ferromagnetic and dense paramagnetic phases has shifted to higher temperatures.

\begin{figure}
\begin{center}
\leavevmode
\includegraphics[scale=0.5]{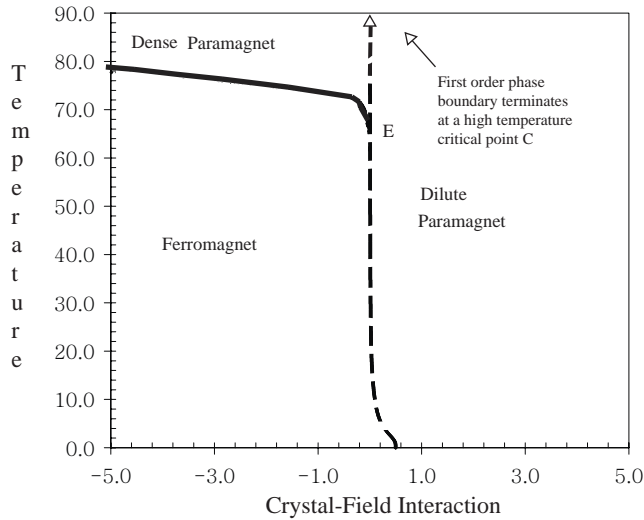} 
\end{center}
\caption{Phase diagram with connectivity values
($p, m_1, m_2, p_A, p_B$) = (8, 8, 9, 40, 1), showing different basins of attraction and \textsl{}associated phases with critical endpoint (E) and critical point (C). Solid lines represent second-order transitions, whereas dashed lines represent first-order transitions.}
\label{fig:Figure 6}
\end{figure}

In our final two phase diagrams we consider the effects of increasing the internal connectivity to p=4 (Figure 5) and p=8 (Figure 6).  Topologically the underlying phase diagrams remain unchanged with this increase in the strength of the internal coupling coefficients.  In each phase diagram investigated, however, there is evidence (at low temperatures) of an ordering transition driven by a decrease in temperature that would result in the dilute paramagnetic phase ordering to the ferromagnetic phase via a first order transition over a very small range of crystal-field interactions.  The shift in the second-order phase boundary separating the dense paramagnetic and ferromagnetic phases continues to be pushed to higher temperatures with increasing internal connectivity.

In each plane in parameter space considered in this study, lines of critical transitions have been probed.  Linearization of the recursion relations while maintaining scaling fields associated with $J, K, \Delta$, and, nonzero perturbations in odd sector contributions $H$ and $L$, results in a recursion matrix.  Calculation of the eigenvalues allow us to extract critical scaling exponents (as discussed in further detail in Seciton 4.  This analysis has been conducted for the line of criticality separating the ferromagnetic and dense paramagnetic states at high temperatures for each degree of internal connectivity (i.e. $p = 1, 2, 4,$ and $8$).  

Associated with scaling fields $J, K, \Delta$ for the case of $p=1$, we find two relevant eigenvalues  $\Lambda_2 = 13.48$ and $\Lambda_4 = 2.00$; corresponding to critical scaling exponents of $y_2 = 3.75$ and $y_4 = 1.00$, respectively; and, an irrelevant eigenvalue with $\Lambda_6 = -1.48$.  Associated with the odd sector scaling fields, $H$ and $L$, we find two relevant eigenvalues  $\Lambda_1 = 5.00$ and $\Lambda_3 = 2.00$; corresponding to critical scaling exponents of $y_1 = 2.32$ and $y_3 = 1.00$.

For internal connectivity $p=2$, associated with scaling fields $J, K, \Delta$, we find two relevant eigenvalues $\Lambda_2 = 15.54$ and $\Lambda_4 = 2.00$; corresponding to critical scaling exponents of $y_2 = 3.96$ and $y_4 = 1.00$, respectively; and, an irrelevant eigenvalue with $\Lambda_6 = -1.54$.  Associated with $H$ and $L$ we find two relevant eigenvalues  $\Lambda_1 = 6.00$ and $\Lambda_3 = 2.00$; corresponding to critical scaling exponents of $y_1 = 2.58$ and $y_3 = 1.00$.

For the case of $p=4$, associated with $J, K, $ and $\Delta$, we find two relevant eigenvalues $\Lambda_2 = 19.63$ and $\Lambda_4 = 2.00$; corresponding to critical scaling exponents of $y_2 = 4.29$ and $y_4 = 1.00$, respectively; and, an irrelevant eigenvalue with $\Lambda_6 = -1.63$.  Associated with $H$ and $L$ we find $\Lambda_1 = 8.00$ and $\Lambda_3 = 2.00$; corresponding to critical scaling exponents of $y_1 = 3.00$ and $y_3 = 1.00$.

In the last plane in parameter space, corresponding to $p = 8$, associated with scaling fields $J, K, \Delta$, we find two relevant eigenvalues $\Lambda_2 = 27.73$ and $\Lambda_4 = 2.00$; corresponding to critical scaling exponents of $y_2 = 4.79$ and $y_4 = 1.00$, respectively; and, an irrelevant eigenvalue with $\Lambda_6 = -1.73$.  Associated with the odd sector fields, $H$ and $L$, we find two relevant eigenvalues  $\Lambda_1 = 12.00$ and $\Lambda_3 = 2.00$; corresponding to critical scaling exponents of $y_1 = 3.58$ and $y_3 = 1.00$.

\section{Summary}
In summary, this study considers a Blume-Capel Ising ferromagnet using renormalization group methods and a hierarchical lattice.  Exact recursion relations are developed, phase diagrams calculated and critical exponents extracted as the concentration of annealed vacancies and temperature are varied.

The results presented here detail our investigation into the effects of varying the strength of the internal coupling coefficients via the connectivity parameter p.  Three unique regions were found in each plane considered: dense paramagnetic, dilute paramagnetic and ferromagnetic.  The high temperature dense paramagnetic/ferromagnetic boundary was found to be second-order.  This critical line terminating at the critical endpoint E.  The qualitative topology of each underlying phase diagram remained unchanged with increasing internal connectivity p.  The only subtle change being the gradual increase, with increasing p, of the critical line separating the high temperature dense paramagnetic and ferromagnetic phases.

The primary goal of this work has been, and remains, to develop a complete and better understanding of the effects of varying temperature and density upon ordering and criticality in ferromagnets with a range of internal coupling coefficient strengths.  Previous studies by this author ~\cite{Snowman:JMMM07, Snowman:JMMM08, Snowman:PRE08} have considered the role of various types of competing interactions using the same renormalization-group approach on hierarchical lattices with similar connectivity parameters.  Thus, the present study - complete with uniform ferromagnetic and crystal-field interactions - can also be used as a standard for comparison when probing the effects due to competing bilinear and/or crystal-field interactions.

\section{acknowledgements}
The author would like to thank Rhode Island College for research release time, and institutional resources, in support of this work.

%\bibliography{SnowmanReferences}

\begin{thebibliography}{30}
\expandafter\ifx\csname natexlab\endcsname\relax\def\natexlab#1{#1}\fi
\expandafter\ifx\csname bibnamefont\endcsname\relax
  \def\bibnamefont#1{#1}\fi
\expandafter\ifx\csname bibfnamefont\endcsname\relax
  \def\bibfnamefont#1{#1}\fi
\expandafter\ifx\csname citenamefont\endcsname\relax
  \def\citenamefont#1{#1}\fi
\expandafter\ifx\csname url\endcsname\relax
  \def\url#1{\texttt{#1}}\fi
\expandafter\ifx\csname urlprefix\endcsname\relax\def\urlprefix{URL }\fi
\providecommand{\bibinfo}[2]{#2}
\providecommand{\eprint}[2][]{\url{#2}}

\bibitem[{\citenamefont{{M. Blume} et~al.}(1971)\citenamefont{{M. Blume}, {V.J.
  Emery}, and {R.B. Griffiths}}}]{Blume:PRA71}
\bibinfo{author}{\bibnamefont{{M. Blume}}}, \bibinfo{author}{\bibnamefont{{V.J.
  Emery}}}, \bibnamefont{and} \bibinfo{author}{\bibnamefont{{R.B. Griffiths}}},
  \bibinfo{journal}{Phys. Rev. A} \textbf{\bibinfo{volume}{4}},
  \bibinfo{pages}{1071} (\bibinfo{year}{1971}).

\bibitem[{\citenamefont{{T.R. Kirkpatrick} and {D.
  Thirumalai}}(1987)}]{Kirkpatrick:PRB97}
\bibinfo{author}{\bibnamefont{{T.R. Kirkpatrick}}} \bibnamefont{and}
  \bibinfo{author}{\bibnamefont{{D. Thirumalai}}}, \bibinfo{journal}{Phys. Rev.
  B} \textbf{\bibinfo{volume}{36}}, \bibinfo{pages}{5388}
  (\bibinfo{year}{1987}).

\bibitem[{\citenamefont{{K.E. Newman} and {J. D. Dow}}(1983)}]{Newman:PRB83}
\bibinfo{author}{\bibnamefont{{K.E. Newman}}} \bibnamefont{and}
  \bibinfo{author}{\bibnamefont{{J. D. Dow}}}, \bibinfo{journal}{Phys. Rev. B}
  \textbf{\bibinfo{volume}{27}}, \bibinfo{pages}{7495} (\bibinfo{year}{1983}).

\bibitem[{\citenamefont{Schick and {W.-H. Shih}}(1986)}]{Schick:PRB86}
\bibinfo{author}{\bibfnamefont{M.}~\bibnamefont{Schick}} \bibnamefont{and}
  \bibinfo{author}{\bibnamefont{{W.-H. Shih}}}, \bibinfo{journal}{Phys. Rev. B}
  \textbf{\bibinfo{volume}{34}}, \bibinfo{pages}{1797} (\bibinfo{year}{1986}).

\bibitem[{\citenamefont{Coniglio}(1993)}]{Coniglio:JPF93}
\bibinfo{author}{\bibfnamefont{A.}~\bibnamefont{Coniglio}},
  \bibinfo{journal}{J. Phys. IV France} \textbf{\bibinfo{volume}{3}},
  \bibinfo{pages}{C1} (\bibinfo{year}{1993}).

\bibitem[{\citenamefont{Maritan et~al.}(1992)\citenamefont{Maritan, Cieplak,
  {M. R. Swift}, and Toigo}}]{Maritan:PRL92}
\bibinfo{author}{\bibfnamefont{A.}~\bibnamefont{Maritan}},
  \bibinfo{author}{\bibfnamefont{M.}~\bibnamefont{Cieplak}},
  \bibinfo{author}{\bibnamefont{{M. R. Swift}}}, \bibnamefont{and}
  \bibinfo{author}{\bibfnamefont{F.}~\bibnamefont{Toigo}},
  \bibinfo{journal}{Phys. Rev. Lett.} \textbf{\bibinfo{volume}{69}},
  \bibinfo{pages}{221} (\bibinfo{year}{1992}).

\bibitem[{\citenamefont{{W. Hoston} and {A.N. Berker}}(1991)}]{Hoston:PRL91}
\bibinfo{author}{\bibnamefont{{W. Hoston}}} \bibnamefont{and}
  \bibinfo{author}{\bibnamefont{{A.N. Berker}}}, \bibinfo{journal}{Phys. Rev.
  Lett} \textbf{\bibinfo{volume}{67}}, \bibinfo{pages}{1027}
  (\bibinfo{year}{1991}).

\bibitem[{\citenamefont{Sellitto et~al.}(1997)\citenamefont{Sellitto, Nicodemi,
  and {J.J. Arenzon}}}]{Sellitto:JPF97}
\bibinfo{author}{\bibfnamefont{M.}~\bibnamefont{Sellitto}},
  \bibinfo{author}{\bibfnamefont{M.}~\bibnamefont{Nicodemi}}, \bibnamefont{and}
  \bibinfo{author}{\bibnamefont{{J.J. Arenzon}}}, \bibinfo{journal}{J. Phys. I
  France} \textbf{\bibinfo{volume}{7}}, \bibinfo{pages}{945}
  (\bibinfo{year}{1997}).

\bibitem[{\citenamefont{{D. P. Snowman}}(1995)}]{Snowman:PhD95}
\bibinfo{author}{\bibnamefont{{D. P. Snowman}}}, \bibinfo{type}{{PhD}
  dissertation}, \bibinfo{school}{University of Maine},
  \bibinfo{address}{Orono, ME} (\bibinfo{year}{1995}).

\bibitem[{\citenamefont{{S.R. McKay} and {A.N. Berker}}(1984)}]{McKay:JAP84}
\bibinfo{author}{\bibnamefont{{S.R. McKay}}} \bibnamefont{and}
  \bibinfo{author}{\bibnamefont{{A.N. Berker}}}, \bibinfo{journal}{J. Appl.
  Phys.} \textbf{\bibinfo{volume}{55}}, \bibinfo{pages}{1646}
  (\bibinfo{year}{1984}).

\bibitem[{\citenamefont{{A.N. Berker} and {M. Wortis}}(1976)}]{Berker:PRB76}
\bibinfo{author}{\bibnamefont{{A.N. Berker}}} \bibnamefont{and}
  \bibinfo{author}{\bibnamefont{{M. Wortis}}}, \bibinfo{journal}{Phys. Rev. B}
  \textbf{\bibinfo{volume}{14}}, \bibinfo{pages}{4946} (\bibinfo{year}{1976}).

\bibitem[{\citenamefont{{S.R. McKay} and {A.N. Berker}}(1982)}]{McKay:PRL82}
\bibinfo{author}{\bibnamefont{{S.R. McKay}}} \bibnamefont{and}
  \bibinfo{author}{\bibnamefont{{A.N. Berker}}}, \bibinfo{journal}{Phys. Rev.
  Lett.} \textbf{\bibinfo{volume}{48}}, \bibinfo{pages}{767}
  (\bibinfo{year}{1982}).

\bibitem[{\citenamefont{{N. S. Branco}}({\natexlab{a}})}]{Branco:lanl99}
\bibinfo{author}{\bibnamefont{{N. S. Branco}}},
  \bibinfo{note}{arXiv:cond-mat/9904082v1}.

\bibitem[{\citenamefont{{N. S. Branco} and {B. M Boechat}}()}]{Branco:lanl97}
\bibinfo{author}{\bibnamefont{{N. S. Branco}}} \bibnamefont{and}
  \bibinfo{author}{\bibnamefont{{B. M Boechat}}},
  \bibinfo{note}{arXiv:cond-mat/9708043v3}.

\bibitem[{\citenamefont{{D.P. Snowman}}(2007)}]{Snowman:JMMM07}
\bibinfo{author}{\bibnamefont{{D.P. Snowman}}}, \bibinfo{journal}{J. Magn.
  Magn. Mater.} \textbf{\bibinfo{volume}{314}}, \bibinfo{pages}{69}
  (\bibinfo{year}{2007}).

\bibitem[{\citenamefont{{D.P. Snowman}}(2008{\natexlab{a}})}]{Snowman:JMMM08}
\bibinfo{author}{\bibnamefont{{D.P. Snowman}}}, \bibinfo{journal}{J. Magn.
  Magn. Mater.} \textbf{\bibinfo{volume}{320}}, \bibinfo{pages}{1622}
  (\bibinfo{year}{2008}{\natexlab{a}}).

\bibitem[{\citenamefont{{D.P. Snowman}}(2008{\natexlab{b}})}]{Snowman:PRE08}
\bibinfo{author}{\bibnamefont{{D.P. Snowman}}}, \bibinfo{journal}{Phys. Rev. E}
  \textbf{\bibinfo{volume}{77}}, \bibinfo{pages}{041112}
  (\bibinfo{year}{2008}{\natexlab{b}}).

\bibitem[{\citenamefont{Falicov and {A.N. Berker}}(1996)}]{Falicov:PRL96}
\bibinfo{author}{\bibfnamefont{A.}~\bibnamefont{Falicov}} \bibnamefont{and}
  \bibinfo{author}{\bibnamefont{{A.N. Berker}}}, \bibinfo{journal}{Phys. Rev.
  Lett.} \textbf{\bibinfo{volume}{76}},\bibinfo{pages}{4380} (\bibinfo{year}{1996}).

\bibitem[{\citenamefont{Kabakcioglu and {A.N. Berker}}(1999)}]{Kabak:PRL99}
\bibinfo{author}{\bibfnamefont{A.}~\bibnamefont{Kabakcioglu}} \bibnamefont{and}
  \bibinfo{author}{\bibnamefont{{A.N. Berker}}}, \bibinfo{journal}{Phys. Rev.
  Lett.} \textbf{\bibinfo{volume}{82}}, \bibinfo{pages}{2572} (\bibinfo{year}{1999}).

\bibitem[{\citenamefont{{N. S. Branco}}({\natexlab{b}})}]{Branco:lanl98}
\bibinfo{author}{\bibnamefont{{N. S. Branco}}},
  \bibinfo{note}{arXiv:cond-mat/9803220v1}.

\bibitem[{\citenamefont{{A.N. Berker} and Ostlund}(1979)}]{Berker:JPC79}
\bibinfo{author}{\bibnamefont{{A.N. Berker}}} \bibnamefont{and}
  \bibinfo{author}{\bibfnamefont{S.}~\bibnamefont{Ostlund}},
  \bibinfo{journal}{J. Phys. C} \textbf{\bibinfo{volume}{12}},
  \bibinfo{pages}{4961} (\bibinfo{year}{1979}).

\bibitem[{\citenamefont{Kaufman and {R.B. Griffiths}}(1981)}]{Kauffman:PRB81}
\bibinfo{author}{\bibfnamefont{M.}~\bibnamefont{Kaufman}} \bibnamefont{and}
  \bibinfo{author}{\bibnamefont{{R.B. Griffiths}}}, \bibinfo{journal}{Phys.
  Rev. B} \textbf{\bibinfo{volume}{24}}, \bibinfo{pages}{496}
  (\bibinfo{year}{1981}).

\bibitem[{\citenamefont{Migliorini and {A.N. Berker}}(1998)}]{Migliorini:PRB98}
\bibinfo{author}{\bibfnamefont{G.}~\bibnamefont{Migliorini}} \bibnamefont{and}
  \bibinfo{author}{\bibnamefont{{A.N. Berker}}}, \bibinfo{journal}{Phys. Rev.
  B} \textbf{\bibinfo{volume}{57}}, \bibinfo{pages}{426}
  (\bibinfo{year}{1998}).

\bibitem[{\citenamefont{Andelman and {A.N. Berker}}(1984)}]{Andelman:PRB84}
\bibinfo{author}{\bibfnamefont{D.}~\bibnamefont{Andelman}} \bibnamefont{and}
  \bibinfo{author}{\bibnamefont{{A.N. Berker}}}, \bibinfo{journal}{Phys. Rev.
  B} \textbf{\bibinfo{volume}{29}}, \bibinfo{pages}{2630}
  (\bibinfo{year}{1984}).

\bibitem[{\citenamefont{Falicov et~al.}(1995)\citenamefont{Falicov, {A.N.
  Berker}, and {S.R. McKay}}}]{Falicov:PRB95}
\bibinfo{author}{\bibfnamefont{A.}~\bibnamefont{Falicov}},
  \bibinfo{author}{\bibnamefont{{A.N. Berker}}}, \bibnamefont{and}
  \bibinfo{author}{\bibnamefont{{S.R. McKay}}}, \bibinfo{journal}{Phys. Rev. B}
  \textbf{\bibinfo{volume}{51}}, \bibinfo{pages}{8266} (\bibinfo{year}{1995}).

\bibitem[{\citenamefont{{R.A. da Silveira} and {J.P.
  Bouchaud}}(2004)}]{daSilv:PRL04}
\bibinfo{author}{\bibnamefont{{R.A. da Silveira}}} \bibnamefont{and}
  \bibinfo{author}{\bibnamefont{{J.P. Bouchaud}}}, \bibinfo{journal}{Phys. Rev.
  Lett.} \textbf{\bibinfo{volume}{93}}, \bibinfo{pages}{015901}
  (\bibinfo{year}{2004}).

\bibitem[{\citenamefont{{R.B. Stinchcombe} and {A.C.
  Maggs}}(1986)}]{Stinchcombe:JPA86}
\bibinfo{author}{\bibnamefont{{R.B. Stinchcombe}}} \bibnamefont{and}
  \bibinfo{author}{\bibnamefont{{A.C. Maggs}}}, \bibinfo{journal}{J. Phys. A}
  \textbf{\bibinfo{volume}{19}}, \bibinfo{pages}{1949} (\bibinfo{year}{1986}).

\bibitem[{\citenamefont{{S.R. McKay} et~al.}(1982)\citenamefont{{S.R. McKay},
  {A.N. Berker}, and {S. Kirkpatrick}}}]{McKay:JAP82}
\bibinfo{author}{\bibnamefont{{S.R. McKay}}},
  \bibinfo{author}{\bibnamefont{{A.N. Berker}}}, \bibnamefont{and}
  \bibinfo{author}{\bibnamefont{{S. Kirkpatrick}}}, \bibinfo{journal}{J. Appl.
  Phys.} \textbf{\bibinfo{volume}{53}}, \bibinfo{pages}{7974}
  (\bibinfo{year}{1982}).

\bibitem[{\citenamefont{{A.A. Migdal}}(1975)}]{Migdal75}
\bibinfo{author}{\bibnamefont{{A.A. Migdal}}}, \bibinfo{journal}{Zh. Eksp.
  Teor. Fiz} \textbf{\bibinfo{volume}{69}}, \bibinfo{pages}{1457}
  (\bibinfo{year}{1975}).

\bibitem[{\citenamefont{{L.P. Kadanoff}}(1976)}]{Kadanoff76}
\bibinfo{author}{\bibnamefont{{L.P. Kadanoff}}}, \bibinfo{journal}{Ann Phys.
  New York} \textbf{\bibinfo{volume}{100}}, \bibinfo{pages}{359}
  (\bibinfo{year}{1976}).

\end{thebibliography}

\end{document}